\documentstyle[epsf]{aa}
\begin{document}
\def\cd{cd$^{-1}$}
\def\cds{cd$^{-1}$\,}
\def\kms{km~s$^{-1}$}
\def\kmss{km~s$^{-1}$\,}
   \thesaurus{06(03.13.2; 03.20.8; 08.09.2:HD 2724; 08.15.1; 08.22.2)}
     % A&A Section 6: Form. struct. and evolut. of stars
%
   \title{Line profile analysis of the $\delta$ Scuti star 
HD~2724$\equiv$BB Phe: mode identification and amplitude variations
\thanks{Based on observations collected at the
Coud\'e Auxiliary Telescope of the European
Southern Observatory -- La Silla, Chile (Proposal 60.E-0113)}}

   \author{L. Mantegazza \and E. Poretti}

   \offprints{L. Mantegazza}

   \institute{Osservatorio Astronomico di Brera --
              Via E. Bianchi, 46 -- I-23807 Merate (Italy)\\
             E-mail: luciano@merate.mi.astro.it, poretti@merate.mi.astro.it}

\date{}
\maketitle
\markboth{L. Mantegazza and E. Poretti}{Line profile variations of HD 2724}
\begin{abstract}
The line profile variations of the $\delta$ Scuti star HD 2724$\equiv$BB Phe
were studied on the basis of new 189 high--resolution spectrograms 
covering 52 hours of observations on a baseline of 8.3 days.
By combining these results with those of a previous campaign
13 pulsation modes were identified: 5 of them are both photometric and
spectroscopic, 3 are purely spectroscopic and 5 purely  photometric.
For the first time it was possible to compare spectroscopic data taken in
two different seasons: 6 modes were found to be common to both datasets 
and furthermore strong amplitude variations of the excited modes were detected.
The fit of the line profile variations with a model of non-radial pulsating
star allowed us to obtain a reasonable estimate of the inclination of the
 rotational axis and to propose the $\ell,m$ typing 
of the spectroscopic modes.
The frequency content resembles that of 4 CVn, a $\delta$ Sct star with
similar physical parameters.
\keywords{methods: data analysis -- techniques: spectroscopic -- stars:
individual: HD 2724 -- stars: oscillations -- stars: $\delta$ Sct}
\end{abstract}

%
%________________________________________________________________

\section{Introduction}
The light and line profile variations of the $\delta$ Scuti star HD~2724$\equiv$BB
Phe have been recently studied by Bossi et al. (1998, hereinafter Paper I). 
On the basis
of 11 consecutive nights of photometric observations and 5
simultaneous consecutive nights of spectroscopic ones they detected
13 probable pulsation modes, 7 of which were determined in an unambiguous
way.  For 4 modes it was possible to suggest an identification of their
$\ell$ and $m$ parameters; in particular, the proposed identification of the
 lowest frequency mode as the radial fundamental one allowed a more accurate
refinement of the stellar physical parameters which were also consistent
with the Hipparcos parallax.

Due to the relatively short spectroscopic baseline, some of the modes
detected in Paper I
had barely resolved frequencies; as a consequence, some ambiguities
arose in their detection and in the successive attempts of their typing.
With the aim to confirm and eventually improve these findings,
an application for a longer run was submitted to ESO. In this paper
we describe the results we obtained; unfortunately, in the meantime
all the facilities
for getting photometric data were dismissed and  we had  to
limit ourselves to a purely spectroscopic campaign, to which 10 consecutive
nights were allotted (October 1--11, 1997).  

\section{New spectroscopic observations and data processing}

The spectroscopic observations were made  
 at La Silla Observatory (ESO) with the Coud\'e Echelle Spectrometer
attached to the Coud\'e Auxiliary Telescope. Owing to bad weather, it was
 possible to get useful observations during 8 nights only.
The run  was performed in Remote Control Mode from Garching headquarters; the 
CES was configured in the blue path 
with the long camera and the CCD \#38. The resulting  reciprocal dispersion
was 0.018 \AA~pix$^{-1}$ with an effective resolution of about 54000. 
The useful
spectral region ranges from 4482 to 4534 \AA. The integration time
was set to 15 minutes; a total of  189 useful spectrograms were 
collected covering about 52 hours of stellar monitoring 
on a baseline of 8.3 days.
Data reduction was performed using the MIDAS package.
The spectrograms were normalized by means of internal quartz lamp
flat fields and calibrated into wavelengths by means of a thorium lamp.

Due to stellar projected rotational velocity,
only the Fe{\sc ii} line at 4508.4 \AA~is completely 
free from blends of
adjacent features and allows a good normalization to the stellar continuum.
Therefore, in the same way as we did in Paper~I, we studied the 
behaviour of this line. 
All the spectrograms  were averaged to obtain a very high $S/N$
average spectrum. It was then possible to select two windows
on both sides of the Fe{\sc ii} line. As a further step,
the individual spectrograms were normalized to the continuum defined 
by a linear least--squares fit of these windows. 
Finally, the spectra were shifted and rebinned in order to
remove the observer's motion. In the rebinning procedure the spectrograms
were resampled with a step of 0.04 \AA~ (average of 2 original pixels):
in such a way we saved the effective resolution
according to the Nyquist criterion and we improved the $S/N$ 
of the resulting profiles.
The mean standard deviation of the pixels on the stellar continuum allowed
us to estimate  the $S/N$ of the spectrograms: the resulting 
average value at the continuum level is 368.

A non--linear least--squares fit of a rotationally broadened gaussian
profile to the average line profile  allowed us to estimate 
 the projected rotational velocity and intrinsic
width:  $v\sin i= 83.0\pm1.0$ \kmss and $W_i=13.7\pm0.5~$\kms.
This $v\sin i$ is in excellent agreement with the value of $82\pm 2$ \kmss 
derived  in Paper I.
Figure 1 shows the average profile obtained from the 1997 
observations fitted with the computed rotationally broadened profile (dashed
line).
\begin{figure}[t]
\epsfxsize=8.5truecm
\centerline{\epsffile{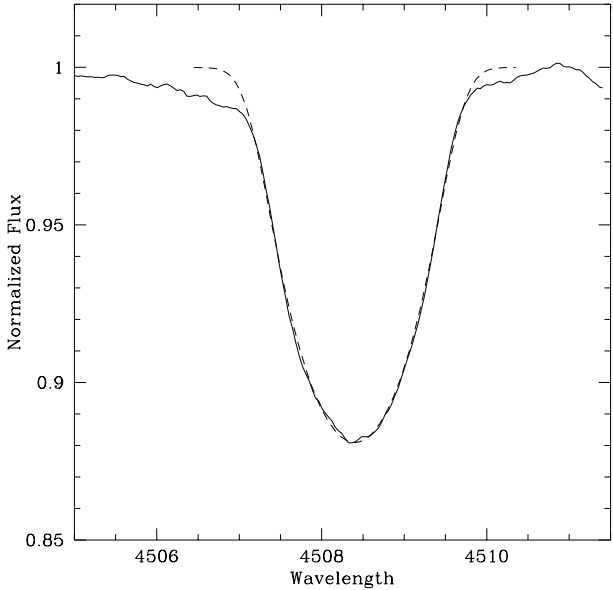}}
\caption[]{Average profile of the Fe{\sc ii} 4508\AA~line with  
the best fitting rotationally broadened profile  superimposed 
($v\sin~i=83.0$~\kms).}
\end{figure}

\section{Analysis of line profile variations}

\subsection{The least--squares algorithm}
The search for periodicities in the line profile variations was performed
by using a generalized form of the least--squares power spectrum technique,
originally developed for the study of monodimensional time series analysis by
Vani\^cek (1971). Let $P(\lambda_j,t_k)$ the observed line
profiles, $j$ the pixel number and $t_k$ the time of the $k$-th 
spectrogram. The global variance is defined by:  
$$\sigma_T=\sum_{j,k}w_k^2(P(\lambda_j,t_k)-P_0(\lambda_j))^2$$ 
where $P_0(\lambda_j)$ is the time averaged profile and $w_k$ are the normalized 
weights derived from the $S/N$ of the spectrograms.
If we have already detected $m$ periodic sinusoidal components (``known
constituents'') and if we are looking for the $(m+1)$-th component, we can
explore the useful frequency range (0$<\nu_i<25$ \cd) by fitting 
each pixel time series $j$ with the series 
\begin{eqnarray}
 p_{i,j}(t_k)=\overline p_{i}+\sum_{l=1,m} A_{i,j,l} \cos(2\pi\nu_l t_k
+\phi_{i,j,l}) \nonumber \\
+ A_{i,j,m+1} \cos(2\pi\nu_i t_k+\phi_{i,j,m+1}) \nonumber 
\end{eqnarray}
where $\overline p_{i},A_{i,j,l}$ and $\phi_{i,j,l}$ (with $1\leq l\leq m+1$)
 are the free parameters. 
Then we compute the global reduction of variance defined as
$$RF_i=1-\sum_{j,k}w_k^2(p_{i,j}(t_k)-P(\lambda_j,t_k))^2/\sigma_m $$ 
where $\sigma_m$ is the global residual variance after the fit of the line 
profile variations with the $m$ ``known constituents''.
The frequency $\nu_i$ giving the highest $RF$ 
(or one of its 1~\cds aliases if there is a better agreement with the 
photometrically detected modes) is then selected as the
$(m+1)$--th known constituent and the procedure is iterated again. 
At the end of this procedure, after having detected $M$ known constituents,
we can perform a final fit of $P(\lambda_j,t_k)$ and derive the functions:
$\overline p_M(\lambda_j)$ (i.e. the best estimate of the unperturbed
line profile), $A_l(\lambda_j), \phi_l(\lambda_j)$ (with $1\leq l\leq M$)
 and also  their formal errors.

\subsection{Frequency identification in 1997 data}
Following the procedure described above,
8 terms have been successively detected:
8.58, 8.05, 6.49, 0.07 (or 1.07), 5.74, 5.31, 7.38 and 6.99 \cd.
Other smaller amplitude terms are probably present, but we cannot assign
a frequency value in a reliable way.
The detected modes are listed in Tab.~1 in order of increasing frequency
(first column); the second column reports their rms amplitudes
computed along the whole line profile and expressed in thousandths of the 
continuum amplitude.

\begin{table}
\begin{flushleft}
\caption{Amplitudes of the photometric and spectroscopic modes; values
between brackets are uncertain (see text). The
amplitude of the 6.99 \cds term cannot be determined in the 1993 data owing
to the low frequency resolution. Spectroscopic amplitudes are in units of
the continuum intensity.}
\begin{tabular}{r c cc r c}
\hline
\noalign{\smallskip}
& & \multicolumn{2}{c}{Spectroscopy}& &
\multicolumn{1}{c}{Photometry}\\
\multicolumn{1}{c}{Freq.} & & \multicolumn{1}{c}{1997} &
\multicolumn{1}{c}{1993} & & \multicolumn{1}{c}{$b$ light}\\
\multicolumn{1}{c}{[\cd]}& &\multicolumn{1}{c}{[10$^{-3}$]} &
\multicolumn{1}{c}{[10$^{-3}$]} & & \multicolumn{1}{c}{[mmag]}\\
\noalign{\smallskip}
\hline          
4.43 & & --   & --   & & 8.4 \\
4.54 & & --   & --   & & 2.8 \\
5.31 & & 1.67 & (3.43)& & 6.7 \\ 
5.63 & & --   & --   & & 2.9 \\
5.74 & & 2.14 & 4.93 & & 11.1 \\
5.88 & & --   & --   & & 3.9 \\
6.12 & & --   & --   & & 5.6 \\
6.26 & & --   & (5.22) & & -- \\
6.49 & & 1.88 & 2.95 & & 4.8 \\
6.99 & & 2.15 & --   & & -- \\
7.38 & & 2.10 & 2.44 & & 9.1 \\
8.05 & & 2.73 & 3.55 & & 3.6 \\
8.58 & & 2.78 & 2.40 & & -- \\
\noalign{\smallskip}
\hline
\end{tabular}
\end{flushleft}
\end{table}

Frequency detection by means of the least--squares technique can
seem a subjective procedure when dealing with single--site observations.
As our experience on photometric data proves, this is not true: as an
example, see the frequencies detected by Mantegazza et al.
(1994) on the basis of single--site observations and their confirmation
by Breger et al. (1995) on the basis of multisite ones.
However, to give an independent confirmation of these results, we performed a
second analysis by using the CLEAN algorithm 
generalized so as to analyse line profile variations (Mantegazza et al. 1999,
in preparation).
The averaged spectrum of the line profile variations is shown in  Fig. 2.
\begin{figure}[t]
\epsfxsize=8.5truecm
\centerline{\epsffile{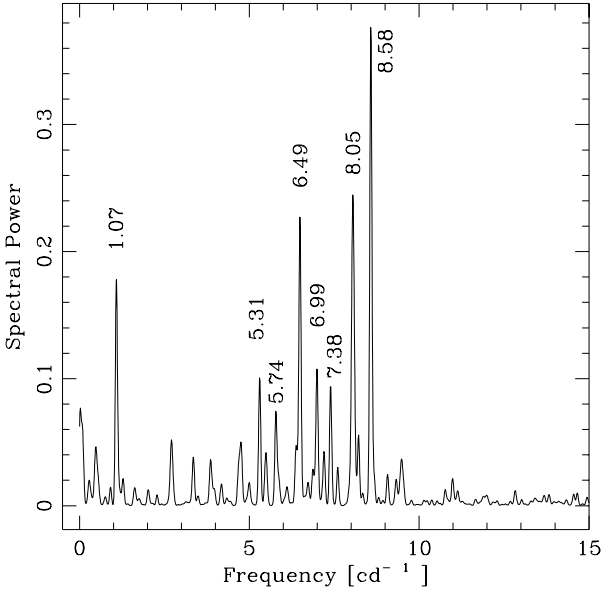}}
\caption[]{CLEAN spectrum of the new data collected in 1997. The
identified peaks are discussed in the text.}
\end{figure}

The only difference between the two algorithms is the value 
assigned to the low frequency term: 0.07 \cds from the least--squares analysis
and 1.07 \cds from the CLEAN algorithm (of course, one term is an alias of the other).
However, we do not assign great significance to this peak: given its closeness
to the value of the sidereal day it is probably an artifact due to 
observation and/or reduction procedures. In particular, during the reductions 
it has been observed
that a slight rotation of the spectrograms on the CCD occurred during
the night; this could be responsible for this low frequency term. 
Therefore, this term will not be considered any more in the following.
All the other terms detected independently by the two techniques are 
coincident. 
\begin{figure*}[]
\epsfxsize=17truecm
\centerline{\epsffile{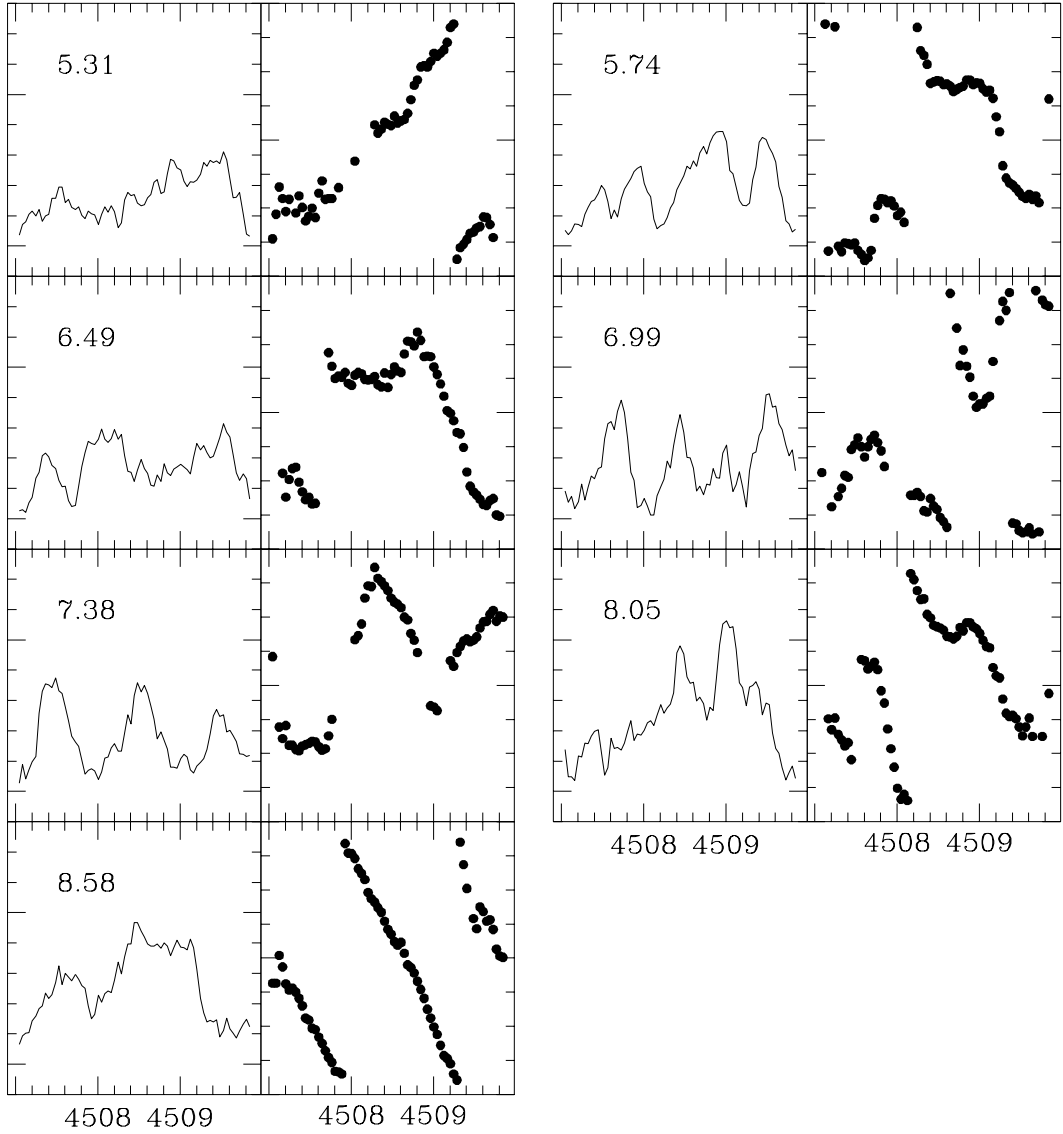}}
\caption[]{Amplitudes and phases of the modes detected by the 
least-squares power spectrum analysis in the 1997 data. All the diagrams
have the same scales: amplitudes are in units of the continuum intensity
(each tick corresponds to 0.001) and phases are in degrees (each tick
corresponds to 50$^o$).}
\end{figure*}
\subsection{Comparison with the 1993 results}
In the third column of Tab.~1 we list the rms
amplitudes of the terms detected from the 1993 spectrograms. 
These values were obtained by including in the least--squares fit both the 5.31
and the 6.26 \cd terms. Owing to the fact that they are barely resolved,
their amplitudes are rather uncertain. Furthermore, 
the amplitudes of the other terms can change by up to about 20\% or so
according to whether one term or both are considered in the fit. 
Notwithstanding these uncertainties  it is quite evident
that, comparing the 1993 and the 1997 data, the amplitudes of the other
terms have decreased,  
with the only exception of the 8.58 \cds term which shows the opposite
behaviour. 
For the modes detected  in the 1993 photometric data,
we have reported their $b$--light amplitude in the fourth column of Tab.~1.

We see that there is a substantial agreement in the frequency detection
between the two seasons: only one mode was detected in 1993 but not in 1997
(6.26 \cd) and only one in 1997 but not in 1993 (6.99 \cd).
The probable reality of the 6.26 \cds term was already discussed in Paper I,
taking also into account its relative closeness to the 1~\cds alias of the
5.31 \cd term.

The 6.99 \cds term was not detected in the 1993 data probably
because it was drowned in the 7.05 \cds alias of the strongest
spectroscopic mode (8.05 \cd). In fact, in 1993 data the peak at 7.05 \cd
was the highest one, probably owing to the 6.99 \cd contribution. 
The correct value (8.05 \cd) was deduced
with the help of the photometry. The resolution of the 1997 data is
sufficient enough to detect 
the 6.99 \cd term: as a matter of fact the least--squares analysis finds 
this term both considering a term at 7.05 or at 8.05 \cd.
 
In summary, five terms were detected in the two
spectroscopic datasets and in the photometric one: 5.736, 5.311, 6.488, 8.049
and 7.382 \cds.
The 6.26, 6.99 and 8.58 \cds are purely spectroscopic terms, while the 4.430,
4.536, 5.629, 5.877 and 6.123 \cds ones are only photometric. 
Even if the detection of these terms in the three different time series
(especially in the spectroscopic ones) was not an easy task, we obtained a
well--defined picture of the pulsational content
of HD 2724 as a final result. In particular, we had for the first time
the possibility
to compare two spectroscopic solutions obtained in different years.

The amplitudes of modes are changed, as it can be seen comparing the different
columns of Tab. 1 where we list the average 
amplitudes along the line profile of the identified modes in the two
observing seasons (cols. 2 and 3).

\subsection{Phase diagrams of spectroscopic terms}

Figure 3 shows the amplitude and phase diagrams of the behaviour along the
line profile of the detected modes (i.e. the functions
$A_l(\lambda_j), \phi_l(\lambda_j), 1\leq l \leq 7$ previously described).
For the sake of clarity only the points with a formal error bar
 smaller than $40^o$ are shown  in the phase panels.

\begin{figure*}[]
\epsfxsize=15.5truecm
\centerline{\epsffile{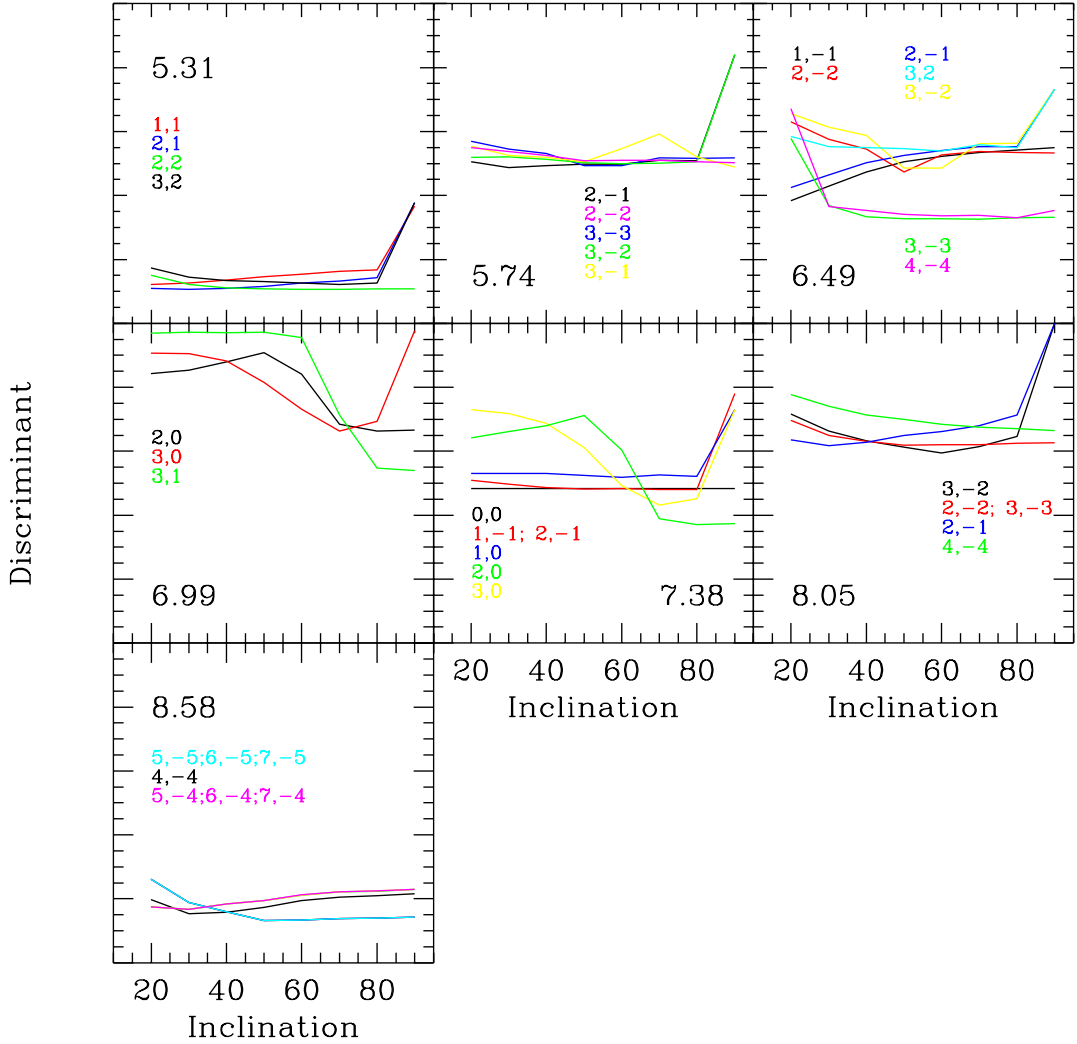}}
\caption[]{Discriminants of the best fitting modes for
the spectroscopically detected terms vs. the inclination of rotational axis
(in degrees). $\ell, m$ numbers are in the same colour of the
corresponding line.}
\end{figure*}
\begin{table}
\begin{flushleft}
\caption{Identification of the pulsation modes. Inclination angle was assumed
to be 70$^o$. With respect to Paper I, the identification of the 5.31
\cds term is changed and the 6.99 \cds term was not previously
detected.}
\begin{tabular}{r c l}
\hline
\noalign{\smallskip}
\multicolumn{1}{c}{Term} & & \multicolumn{1}{c}{Mode typing}\\ 
\multicolumn{1}{c}{[\cd]}& &\multicolumn{1}{c}{($\ell,m$)}\\ 
\noalign{\smallskip}
\hline
5.31 & & $ (2,2)~~ (2,1)~~ (1,1)~~ (3,2) $ \\
5.74 & & $ (2,-2)~~ (2,-1) $ \\
6.49 & & $ (1,-1)~~ (2,-1)~~ (2,-2) $ \\
6.99 & & $ (3,1)~~ (2,0)~~ (3,0)$ \\
7.38 & & $ (1,-1)~~ (2,-1)~~ (0,0)~~ (1,0) $ \\
8.05 & & $ (4,-4)~~ (3,-3)~~ (3,-2) $ \\
8.58 & & $ (5,-5)~~(6,-5)~~(7,-5)$ \\
\noalign{\smallskip}
\hline
\end{tabular}
\end{flushleft}
\end{table}

The behaviours of the phase diagrams are similar for the modes detected 
in the two seasons, with only a notable exception concerning the 5.31
\cds term.
In the analysis of the 1993 data the reality of the 6.26 \cds term was
established, but this term is not very prominent in the 1997 data,
because meanwhile its amplitude may have decreased. However, this dataset
has a better
frequency resolution and the 5.31 and 6.26 \cds terms could be well resolved.
Hence, we can
obtain a more clear phase diagram of the 5.31 \cds term: we verified that
the phase increases with the wavelength,  although we stated 
the opposite behaviour in Paper I. We performed several tests on
the 1993 data and we definitely established that the mode identification
proposed in Paper I suffers from the interference between one term and
the alias at $\pm$ 1 \cds of the other term (in particular the 5.31 \cds term
was affected by the alias at 5.26 \cds of the 6.26 \cds term).
We therefore believe that the most reliable solution should be the one derived
from the new data and that the mode identification proposed in the next
subsection is more reliable than the one  reported in Paper I.

Among the other phase diagrams the cleanest is that of the 8.58 \cds term, 
which clearly indicates that it is a relatively high--degree prograde mode,
in agreement with the results of Paper I. 
\subsection{Mode typing}
An attempt to identify the modes can be accomplished by fitting
the line profile variations with the technique already described in Paper I.
In the case of the present data, 
we are forced to fit the line profile variations only
without the simultaneous light variations;
this makes it difficult to introduce  the flux (or temperature)
variations in the model.
Of course, without this  constraint,  the model obtained from the 1997 data is
a little less well defined, but in spite of that we could obtain important
statements on the pulsational content.
Therefore, we limited ourselves to considering the vertical velocity ($v_r$)
and its phase as the only free parameters of the model, while the horizontal
velocity $v_h$ was kept linked to $v_r$ by the usual relationships 
$v_h=74.4 Q^2 v_r$ ($Q$: pulsational constant) and we assumed the same 
phase for both.
These approximations are legitimated by the fact that $v_h<<v_r$.
Finally, no flux variations were allowed.
The other parameters were the same as in  Paper I, in particular the stellar
physical parameters. 

We explored all the possible combinations of $\ell,m$ up to $\ell=4$ and 
inclination angles between 20$^o$ (at this inclination the rotational
velocity is close to the stellar break--up velocity) and 90$^o$.
For the highest $S/N$ term
(8.58 \cds) term we extended the search to $\ell=7$, because its phase 
diagram clearly indicated that it could be a relatively high--degree mode.

The discriminants of the best fitting modes are reported for each term
in the panels of Fig.~4 as a function of the inclination angle.
It should be noted that in each phase diagram we show only the plausible
solutions, i.e. those giving a low value for the discriminant. The
solutions yielding unacceptable values are omitted for the sake of clarity.
As an example of the quality of the fits supplied by this technique,
Fig.~5 shows the variations of the line profile due to the 8.58 \cds
term phased on a complete cycle and the corresponding best fitting variations
produced by a model of a nonradial mode with $\ell=6, m=-5$ and assuming 
$i=70^o$. As can be seen from Fig.~4 we can be reasonably sure that
the 8.58 \cds is a mode with $\ell=6\pm 1$ and $m=-5$: unfortunately,
we cannot be more precise since the discriminants
of these three modes are practically coincident.  

By examining Fig.~4 we see that the results are not very clear in the sense 
that for most
of the terms different modes give equivalent fits at different inclinations.
However, the 6.99 \cds mode seems to be slightly in favour of
 a solution
with a high inclination (say $i\geq 60^o$). In Paper I we left an
uncertainty on two possible values for the inclination angle, i.e.  
$50^o$ or $70^o$. On the basis of the new  results,  we prefer the latter;
however, most of the proposed mode identifications are plausible also
for the former value.

Owing to the fact that more than one identification is possible for the same
term, the identifications suggested in Paper~I are generally agreed on.  
The strongest disagreement is about the 5.31 \cds term, which should be
prograde or axissymmetric according to the old data while the new ones
indicates clearly that it is retrograde. The reason for this discrepancy
has been already discussed when dealing with the phase diagrams. According
to that discussion we have to be more confident in the results
obtained from 1997 data. 

We report in Tab.~2 the identifications which are compatible both with
the present data and with the 1993 ones. In particular, some possible
identifications suggested by Fig.~4 have
been rejected (as for example the 3,--3 and 4,--4 couples for the 6.49
\cd) because they are unable to fit the light variations observed in
1993.

As regards the 8.05 \cds term, our calculations show that the local
flux of a (4,-4) mode varies of about 0.3 mag and that of a (3,-3)
mode of about 0.15 mag. The cancellation effect reduces it to the
observed mmag level.

\begin{figure}[b]
\epsfxsize=8.8truecm
\centerline{\epsffile{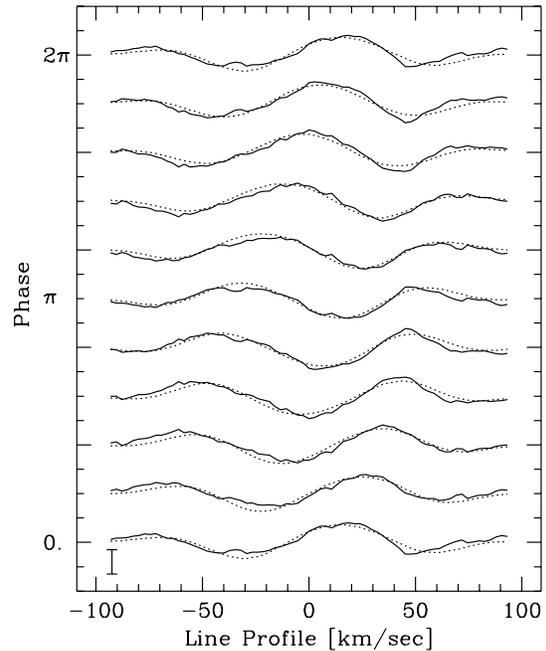}}
\caption[]{The variations induced on the line profile by the 8.58 \cd term
phased over 1 cycle (continuous line) and the best--fitting model of a mode
with $\ell=6$,$m=-5$ and $i=70^o$ (dashed lines). The small bar at the lower left
indicates a 0.005 amplitude  in continuum intensity units.}
\end{figure}
\section{Conclusions} 
The new set of spectrograms of HD~2724 has allowed to confirm the detection
of the modes found in the 1993 data  and  a new mode has been
detected at 6.99 \cd. There is evidence that the amplitudes of
the modes have considerably changed, in particular the 8.58 \cds term,
which was among the weakest in 1993, is now the strongest. Amplitude variations are well
established in $\delta$ Sct stars  thanks to extensive photometric
studies.  A a matter of fact, HD~2724 is
the first star in which these variations are also observed spectroscopically.
In particular, it should be emphasized that the 6.26 \cds term, clearly
detected in the 1993 data, was not recovered in the 1997 ones.

As regards the frequency content, HD 2724 is very similar to 4 CVn
(Breger \& Hiesberger 1999). At least 7 frequencies (5.88,
6.12, 6.26, 6.49, 6.99, 7.38 and 8.58 \cds in HD 2724; 5.85, 6.12,
6.19, 6.44, 6.98, 7.38 and 8.59 \cds in 4 CVn) have almost the same value.
Among the high--amplitude terms in 4 CVn, only the 5.05 \cds term has
no correspondance in HD 2724.
The physical parameters of the two stars are similar:
T$_{\rm eff}$=6900$\pm$100 K, $\log
g$=3.4$\pm$0.1, $v\sin i$=73 \kmss for 4 CVn (Breger et al. 1990),
7200$\pm$100  K, 3.44$\pm$0.03, 83 \kmss for HD 2724 (Paper I). HD 2724
does not show the cross--coupling terms found in the 4 CVn light curve,
but this can be due to the smaller amplitude of the modes excited in HD 2724.
On the basis of this similarity, the comparison between the frequency
content of $\delta$ Sct stars deserves further attention in the future.

The mode typing has been partially hampered by the absence of the light
curve, which prevented us from modelizing flux variations in the line profiles.
%Moreover in order to disentangle the profile variations induced by the
%different modes and thus to get reliable fits, it should be necessary
%to have spectroscopic observations spanning on longer baselines 
%(comparable to those used for photometric data) and possibly also
%obtained with multisite campaigns.
Within these limitations we nevertheless obtained a rather satisfactory
fit of the strongest mode (8.58 \cd) which resulted to have $\ell=6\pm1$
and $m=-5$. 
From the fit to the line profile variations induced by the other
modes some indications of their $\ell,m$ value have been obtained and they
are in general agreement with those of Paper I. A remarkable
result is the retrograde nature of the 5.31 \cds term. Moreover, the
5.74 \cds term have probably $\ell=2$ and $m=-2$ considering the 1993
and 1997 phase diagrams.

The importance of the study of the line profile variations is emphasized
by the detection of modes not observed photometrically. In order to
propose asteroseismological models of $\delta$ Sct stars, the
combination of the two techniques is strongly recommended. It should be
noted that we could obtain 13 independent frequencies in the case of
HD~2724 by single--site campaigns. However, 
 we believe that the results obtained from the 1997 data are very close
to the best effort we could obtain from one site spectroscopic campaigns
lasting 7--10 days. 
\begin{acknowledgements} The authors wish to thank M.~Breger for drawing
their attention to the similarity between HD~2724 and 4~CVn; J.~Vialle 
improved the English form of the manuscript.
\end{acknowledgements}

\end{document}